\documentclass[modern]{aastex631}

\usepackage{amsmath, amsthm, amssymb, amsfonts} 
\usepackage{physics}   
\usepackage{wasysym}

\submitjournal{ApJ}

\begin{document}

\title[SOFIA Observations of the dust-forming WC+O binary WR\,137]{FORCASTing the spectroscopic dust properties of the WC+O binary WR137 with SOFIA}

\author{Megan J. Peatt}
\affiliation{Department of Physics and Astronomy, Embry-Riddle Aeronautical University, 3700 Willow Creek Rd., Prescott, AZ 86301, USA}

\author[0000-0002-2806-9339]{Noel D. Richardson}
\affiliation{Department of Physics and Astronomy, Embry-Riddle Aeronautical University, 3700 Willow Creek Rd., Prescott, AZ 86301, USA}

\author[0000-0002-8092-980X]{Peredur M. Williams}
\affiliation{Institute for Astronomy, University of Edinburgh, Royal Observatory, Edinburgh EH9 3HJ, United Kingdom}

\author[0000-0003-3682-854X]{Nicole Karnath}
\affiliation{Space Science Institute, 4765 Walnut St, Suite B Boulder, CO 80301, USA}
\affiliation{Center for Astrophysics Harvard \& Smithsonian, Cambridge, MA 02138, USA}
% Nicole Karnath nkarnath@gmail.com

\author[0000-0002-2287-8151]{Victor I. Shenavrin}
\affiliation{Sternberg Astronomical Institute, Moscow State University, Universitetskij pr., 13, Moscow, 119991, Russia}

\author[0000-0003-0778-0321]{Ryan M. Lau}
\affiliation{NSF's NOIRLab, 950 N. Cherry Avenue, Tucson, Arizona 85719, USA}

\author[0000-0002-4333-9755]{Anthony F. J. Moffat}
\affiliation{D\'epartement de physique, Universit\'e de Montr\'eal, Complexe des Sciences, 1375 Avenue Th\'er\`ese-Lavoie-Roux, Montr\'eal, Queb\'ec, H2V 0B3, Canada}

\author[0000-0001-9754-2233]{Gerd Weigelt}
\affiliation{Max Planck Institute for Radio Astronomy, Auf dem H\"ugel 69, 53121 Bonn, Germany}

\begin{abstract}

WR\,137 (HD\,192641) is a binary system consisting of a carbon-rich Wolf-Rayet star and an Oe companion star in a 13-year orbit. Near periastron, the winds of the two stars collide and form carbonaceous dust. We obtained three mid-infrared grism spectra of the system with SOFIA and FORCAST during the last year of SOFIA's operations in July 2021, February 2021, and May 2022 (Cycle 9). Within these spectra, we have identified several wind lines from \ion{He}{1}, \ion{He}{2}, \ion{C}{3}, and \ion{C}{4} that are emitted from the Wolf-Rayet wind as well as a weak emission feature around 6.3--6.4 $\mu$m that may have shifted its peak flux from 6.29 to 6.41$\mu$m through this time period. The weak feature grew as the continuum dust emission grew while the WR emission appeared to decline due to lower contrast with the continuum. Furthermore, we observe that the peak of the feature shifts to redder wavelengths during the observations. We compare this feature to the UIR feature and other emission lines identified in dusty WC binaries. For WR\,137, we speculate that mixing of the winds in the system with the Oe star's disk is important for starting the dust formation and that it is less important as dust formation continues. Previous infrared photometry shows ``mini-eruptions" of dust production which could then be explained with variations of the Oe star disk.

\end{abstract}



\section{Introduction} \label{sec:intro}
%Note for Noel when edits start: early Universe carbon dust: https://arxiv.org/abs/2302.05468

Massive stars have large impacts on their environments through strong stellar winds and terminal supernova explosions. The exact nature of the impacts can be changed through binary companions and their orbital architecture. These stars are usually formed in binary systems, and a majority of the systems are expected to interact during their lifetimes \citep{2012Sci...337..444S, 2013A&A...550A.107S}. Amongst the massive stars, Wolf-Rayet (WR) stars represent highly evolved massive stars which have lost their outer hydrogen envelope from either strong stellar winds or binary interactions. They are helium burning and have strong hydrogen-free stellar winds ($\dot{M}\sim 10^{-5} M_\odot\ {\rm yr}^{-1}$; $v_\infty \gtrsim 2000\ {\rm km\ s}^{-1}$) and are assumed to have gone through rapid mass-loss before becoming a WR star. These stars have highly structured, strong stellar winds with terminal speeds of several thousand km s$^{-1}$. 
From their spectra, the WR stars are classified into either nitrogen-rich (WN) or carbon-rich (WC) depending on if the wind emission is dominated by nitrogen and ionized helium or carbon. 

WR\,137 was one of the first WR stars discovered by \citet{1867CRAS...65..292W} at the Paris Observatory along with WR\,134 and WR\,135 (where they were labeled by their 1850 coordinates, the present names coming from the `Sixth Catalogue' \citep{1981SSRv...28..227V}. The latter authors classified its spectrum as WC7+abs, reflecting its uncertain status as a binary. From moderate-resolution spectra, \citet{Underhill62} deduced that it was a WC7+Be shell star binary but \citet{Massey} and \citet{Moffat} did not find variations in its radial velocities, deducing that WR\,137 was not a (close) binary system. 

Multi-wavelength infrared (IR) photometry of late sub-type WC stars found a few to show emission from carbonaceous dust \citep{1972A&A....20..333A}. Two WR stars, WR\,140 (HD\,193793, \citet{1978MNRAS.185..467W}, WC7+abs) and WR\,48a (\citet{1983A&A...118..301D}, WC9) were observed to show rapid increases in the IR attributed to dust formation. Infrared photometry of WR\,137 in 1978--84 \citep{1985MNRAS.215P..23W} also showed a dust-formation episode, with maximum near 1984.5. Re-examination of IR photometry of WR\,137 in 1970--77 when the emission was fading \citep{1976ApJ...210..137H} suggested that WR\,137 was at that time recovering from an earlier, unobserved, dust formation event, leading \citet{1985MNRAS.215P..23W} to suggest that such eruptions recurred at intervals of about 15~years.

The episodic dust formation resembling that by WR\,140 revived the search for binary motion. \citet{Annuk} demonstrated that the WR\,137 system was a binary by presenting a first RV orbit, adopting a period of 4400 days from the IR data but \citet{1992ApJ...398..636U} did not accept this and advocated a single star model for WR\,137. 

Continued IR photometry \citep{2001MNRAS.324..156W} revealed another dust formation episode peaking in 1997, suggesting a period near 4765$\pm$50 days. \citet{2005MNRAS.360..141L} derived a spectroscopic period of 4766$\pm$66 days from their RV orbit, confirming the binary status of WR\,137. \citet{2016MNRAS.461.4115R} collected $H$-band interferometry with the CHARA Array, which allowed them to separate the two components and support a nearly edge-on inclination despite only one epoch of observations. The two stars contribute nearly equally to the $H$-band flux. They also found a spectral type of WC7pd+O9V using spectroscopic modeling.

The WR\,137 dust cloud was imaged in the IR by \citet{1999ApJ...522..433M} in 1997 and 1998, shortly after IR maximum, using  the {\it Hubble Space Telescope}'s NICMOS instrument in the $H^{\prime}-$ and $K^{\prime}-$bands. They found IR-emitting  clumps near the central binary, and evidence for a stream of dust emanating from the central source with an extent of $\sim0.25\arcsec$ ($\sim4$ pc). This type of structure was confirmed by Lau et al.~(2023) who presented {\it JWST} aperture masking interferometric observations with the NIRISS instrument.

The dust may form in the shocked gas between a WC star and companion O star where their winds collide, such as inferred from the rotating `pinwheel' of heated dust made in the WC9 system WR\,104 (which still requires determination of its orbit) and the canonical episodic dust-maker WR\,140. This system was recently imaged by JWST+MIRI in the mid-infrared with three filters spanning 7.7--21$\mu$m, showing more than 150 years of dust created and sent into interstellar space in concentric arcs \citep{2022NatAs...6.1308L}. Spatially-resolved spectroscopy of the dust indicated two unidentified infrared (UIR) features around 6.4 and 7.8 $\mu$m, as well as a strong [S IV] feature at 10.5$\mu$m.
This greatly extended the ground-based IR imaging of WR\,140's dust by \citet{2002ApJ...567L.137M} and \citet{2009MNRAS.395.1749W}.

A decretion disk around the companion star in WR\,137 was discovered from the observation of double-peaked hydrogen and helium emission lines, making it an O9e star. This was first noted by \citet{Underhill62}, and recently investigated in much more detail by \citet{2020MNRAS.497.4448S}. This may explain the constant component of the continuum polarization \citep{2000A&A...361..273H} and the thinner spiral of dust around WR137 \citep[][Lau et al.~submitted]{2020MNRAS.497.4448S}.

\begin{figure*}[t!]
\includegraphics{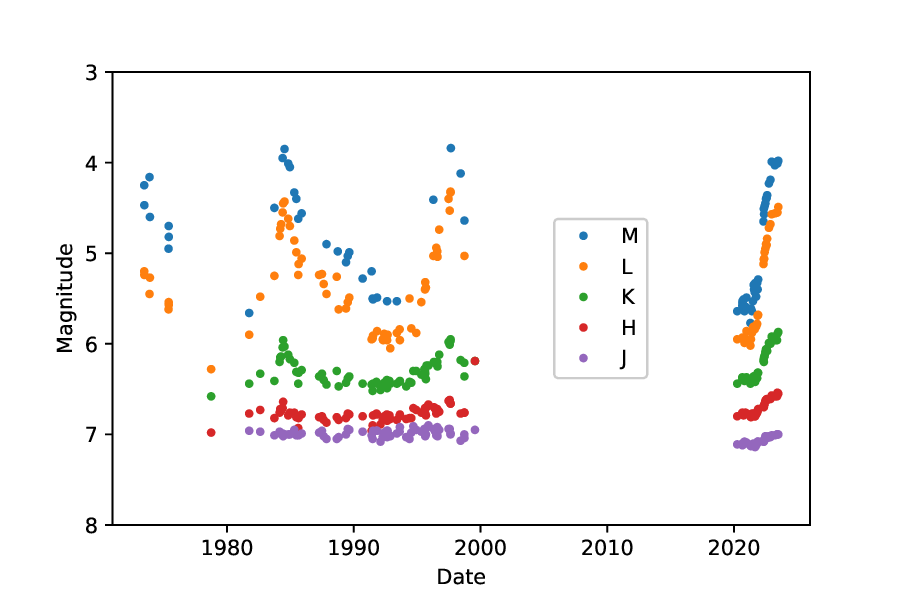}
\caption{The long-term infrared light curve of WR\,137. We show the archival photometry (references in Section 1) of WR\,137 as well as our new measurements. When the binary nears periastron, the infrared brightness increases due to increased dust production with the next periastron happening soon. 
\label{fig:LC}}
\end{figure*}

In this paper, we report on new mid-infrared spectroscopy of WR\,137 taken with the Stratospheric Observatory for Infrared Astronomy (SOFIA) and the FORCAST instrument as the recent dust formation began. We describe these observations in Section 2. Section 3 describes measurements we made of spectral features in our observations, including a measurement of a line that appears to be associated with the dust being created. We discuss these results with a comparison to other WR systems and their dust features in Section 4, concluding this study in Section 5. 
%% outline paper

\section{Observations}

\subsection{Infrared photometry}

Near- and mid-infrared photometry of WR\,137 has been collected at the Sternberg Astronomical Institute’s (SAI) Crimean Laboratory of Moscow State University using a $JHKLM$ photometer with an InSb photovoltaic detector cooled with liquid nitrogen \citep{2011ARep...55...31S}. Observations have been taken regularly since 2020 August, allowing us to compare the outburst with the previously observed outburst. Each data point consists of multiple sub-exposures of 30--60 s with a total integration time of 5–10 minutes in each of the $JHKL$ filters and 20--25 minutes in the $M$ filter. The observations are compared to standard star observations taken before or after each observation of the target. The light curve is tabulated as an online machine readable table. 

%% The "ht!" tells LaTeX to put the figure "here" first, at the "top" next
%% and to override the normal way of calculating a float position

\subsection{SOFIA Observations}

We obtained mid-IR grism spectroscopy using the FORCAST instrument \citep{2013PASP..125.1393H, 2018JAI.....740005H} onboard SOFIA \citep{2018JAI.....740011T}. The first observation, taken in 2021 July, used both the G063 and G111 grating setups, with two subsequent observations in 2022 February and May\footnote{We note that the final flight where our data were collected was nicknamed ``Perry", which is extremely appropriate given Peredur Williams' expertise and involvement in the long-term monitoring and discovery of dust around WC stars.} with only observations using the G063 grism. This setup covers the wavelength range of 4.9–-8.0 $\mu$m, but we found the spectra near the edge of the chip to be too noisy for analysis, effectively reducing the useful range to 5.1--7.9 $\mu$m. These spectra were taken with the 2.4-$\arcsec$ slit, yielding a spectral resolving power of $\sim 180$.
The FORCAST grism data are known to have variable slit-losses, adding uncertainty to the flux calibration, but these data should be accurate to a few percent \citep[e.g.,][]{2021ATel14794....1G}. For this analysis, we focus on the variability of line flux and equivalent widths, so we concentrate on only the G063 spectra.

%% The "ht!" tells LaTeX to put the figure "here" first, at the "top" next
%% and to override the normal way of calculating a float position
\begin{figure*}[ht!]
\includegraphics{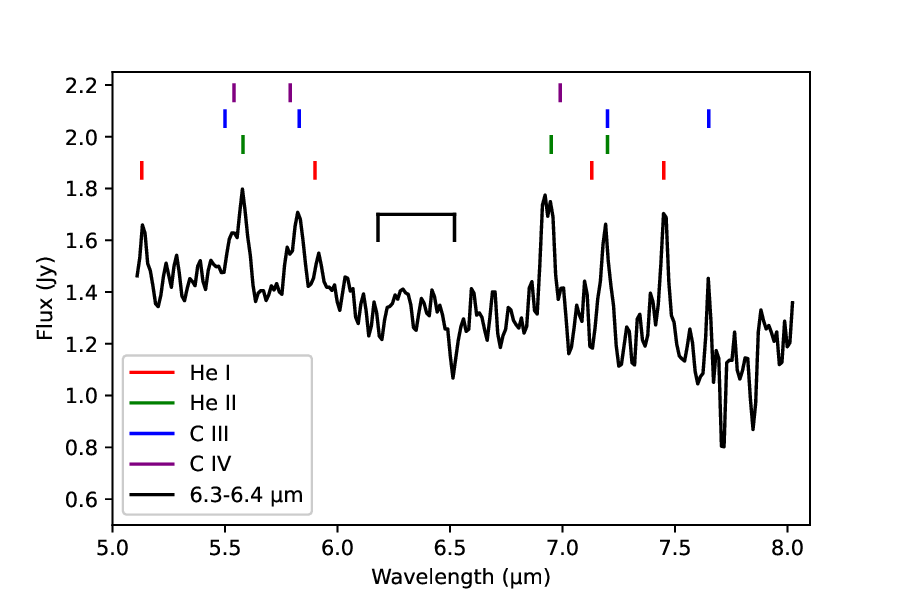}
\includegraphics{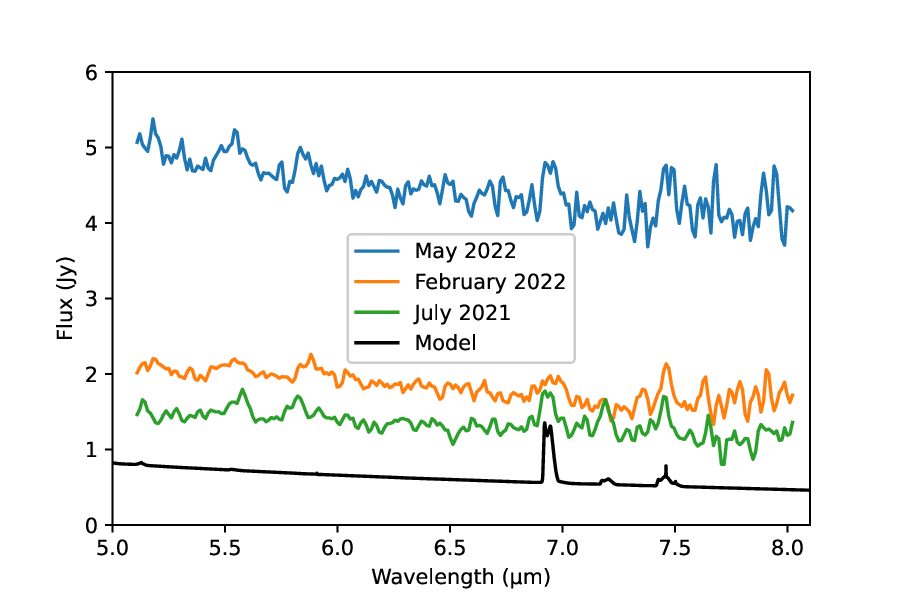}
\caption{In the top panel, we show the spectrum from July 2021 with line identifications for \ion{He}{1}, \ion{He}{2}, \ion{C}{3}, and \ion{C}{4}. We also show the region of the weak UIR feature at 6.3--6.4$\mu$m. In the lower panel, we show the three spectra taken with SOFIA/FORCAST compared to a smoothed model of the binary without dust emission \citep{2016MNRAS.461.4115R}. Note that the model only includes the stellar and wind flux, neglecting all dust emission. Some of these emission lines from the wind were not included in the model.
\label{fig:obs+model}}
\end{figure*}

%% The "ht!" tells LaTeX to put the figure "here" first, at the "top" next
%% and to override the normal way of calculating a float position
\begin{figure*}[ht!]
\includegraphics{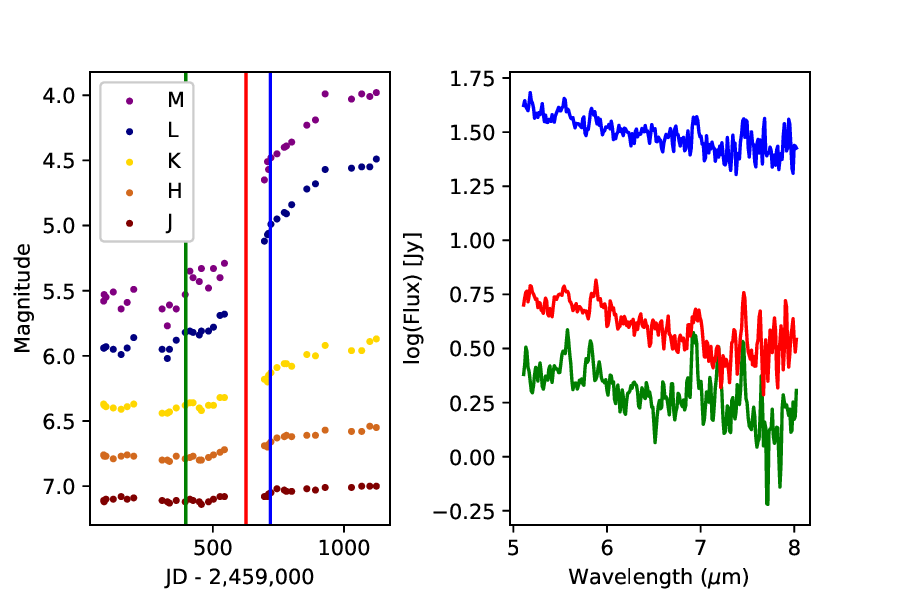}
\caption{Here we show the three spectra with the corresponding times in the infrared evolution. Unfortunately, no IR photometry was able to be collected at times near the February 2022 observation, so the interpolated flux on a line plot is likely overestimated compared with the actual flux at that time. 
\label{fig:LC+spec}}
\end{figure*}

\section{Measurements}

\citet{2016MNRAS.461.4115R} presented models of the combined spectrum of WR\,137 based on optical and ultraviolet spectroscopy. The model, along with the three flux-calibrated spectra from SOFIA, are shown in Fig.~\ref{fig:obs+model}. While the SOFIA data are noisy, a few things are readily apparent. The blended lines at 5.4 and 5.9 $\mu$m are not present in the models presented by \citet{2016MNRAS.461.4115R}. The He I line at $\sim$7.5 $\mu$m appears anomalously strong compared with the model or other helium transitions in our spectra and it is possible that hydrogen Pfund-$\alpha$ (7.46 $\mu$m) from the decretion disk around the companion star reported by \citet{2020MNRAS.497.4448S} could be a significant contributor.
We were able to measure the 6.4 $\mu$m feature equivalent width and then compare it to known WR lines of reasonable strength at 5.6, 5.8, 6.9, and 7.45 $\mu$m. Our errors are statistical in nature as we measured the strength with several assumptions of where the continuum was present, allowing us a reasonable confidence on the strengths of the lines measured.
In Fig.~\ref{fig:obs+model}, we also show line identifications for the spectrum obtained in 2021 July. We identify lines of \ion{He}{1}, \ion{He}{2}, \ion{C}{3}, and \ion{C}{4}, including several that were not included in the PoWR model reported by \citet{2016MNRAS.461.4115R}. 

\begin{figure*}[ht!]
\includegraphics{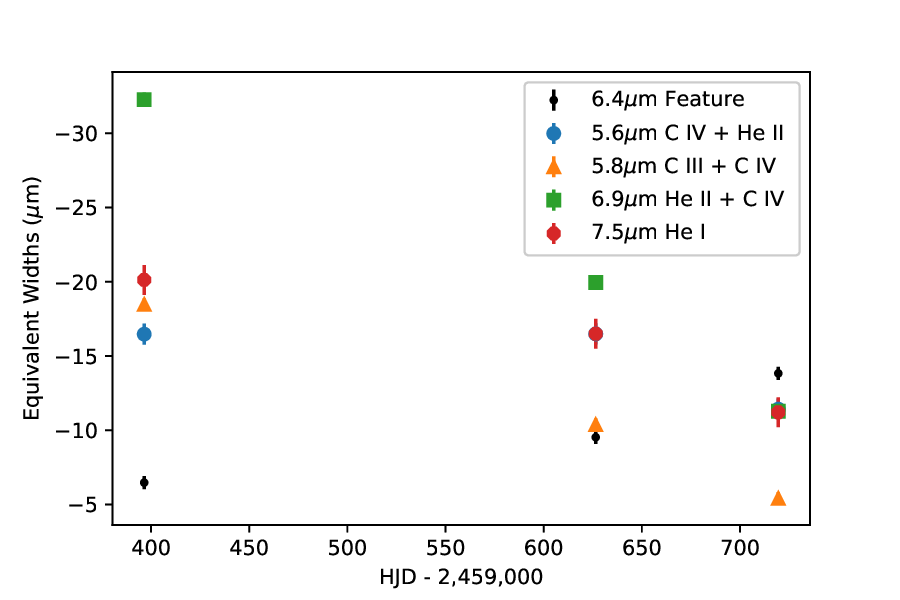}
\caption{The WR lines can be seen decreasing in strength as the continuum increases, while the 6.4$\mu$m feature's strength grows, demonstrating a relation to the dust being produced during periastron. The error bars are the standard deviation from the mean of the measurements. 6.4$\mu$m is shown with a black dot and the others are colored symbols, which match the symbols used in Fig \ref{fig:EWratio}.
\label{fig:EW-time}}
\end{figure*}

The spectra are compared with the IR light curve in Fig.~\ref{fig:LC+spec}, showing that the changing flux observed by SOFIA was seen in the NIR photometry. We show our equivalent width measurements for the 6.4$\mu$m feature as well as other lines in Fig.~\ref{fig:EW-time}. To the first order, we see that the equivalent width of the lines from the wind appear to decrease as the continuum increased through these epochs. This is typical behavior for the wind lines given the additional continuum flux from dust being formed in the wind, as demonstrated in the infrared photometry (Fig.~\ref{fig:LC}). We also show in Table \ref{table} the net flux for the blended lines at 6.9 $\mu$m (the closest line to the UIR band) that shows a constant flux (within uncertainties) across the three observations. 

\begin{figure*}[ht!]
\includegraphics{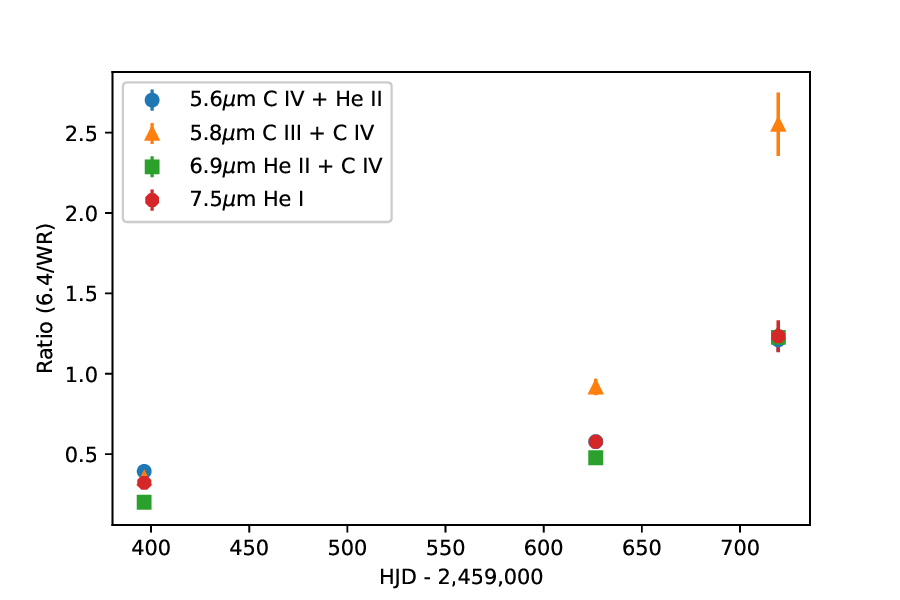}
\caption{To compare the strengths of the lines, a ratio was taken of the 6.4$\mu$m feature over each WR line. Ratios are represented by the corresponding symbol from our previous plot. A clear trend is shown of the ratios steadily increasing over time, though the WR line at 5.8$\mu$m does so at a quicker rate, potentially related to colliding winds or the dust.  
\label{fig:EWratio}}
\end{figure*}

While the equivalent width of the WR wind lines decreased during our three observations, the opposite is true for the 6.4 $\mu$m feature which grew stronger with time. This shows that this feature is inconsistent with the WR wind lines and is therefore associated with the dust production near periastron. We tested this by taking ratios of the 6.4$\mu$m line to each of the wind lines. These ratios, shown in Fig.~\ref{fig:EWratio}, show that this line indeed appears stronger with time, consistent with this feature being formed in the newly formed dust around the system.

\begin{table}
\begin{center}
\begin{tabular}{c  c  c  c} 
\hline
 Line ($\mu$m) & July 2021 & February 2022 & May 2022  \\ 
JD:  & 2,459,396.5 & 2,459,626.5 & 2,459,719.5 \\
    & $W_\lambda$($\mu$m) & $W_\lambda$($\mu$m) & $W_\lambda$($\mu$m) \\
 \hline\hline
 6.3--6.4$\mu$m feature & -6.47 $\pm$ 0.44 [6.29$\mu$m] & -9.53 $\pm$ 0.44 [6.36$\mu$m] & -13.81 $\pm$ 0.44 [6.41$\mu$m] \\ 
 5.6 \ion{C}{4}+\ion{He}{2} & -16.47 $\pm$ 0.71 & -16.48 $\pm$ 0.71 & -11.42 $\pm$ 0.71 \\
 5.8 \ion{C}{3}+\ion{C}{4} & -18.47 $\pm$ 0.41 & -10.38 $\pm$ 0.41 & -5.42 $\pm$ 0.41 \\
 6.9 \ion{C}{4}+\ion{He}{2} & -32.26 $\pm$ 0.47 & -19.95 $\pm$ 0.47 & -11.28 $\pm$ 0.47 \\
 7.5 \ion{He}{1}+\ion{He}{2} & -20.12 $\pm$ 1.00 & -16.50 $\pm$ 1.00 & -11.21 $\pm$ 1.00 \\ 
 \hline
 Flux   & (Jy)  &   (Jy)    & (Jy) \\ \hline
 6.9 \ion{C}{4}+\ion{He}{2} & (4.44$\pm$0.16)$\times 10^{-2}$ &   (4.68$\pm$0.41)$\times 10^{-2}$   &   (4.77$\pm$0.17)$\times 10^{-2}$   \\ 
 \hline
\end{tabular}
\caption{The averages of each equivalent width (in $\mu$m) for the three SOFIA/FORCAST spectra with uncertainties calculated from standard deviation. We note the JD for each spectrum and the central wavelength of the 6.4 $\mu$m is given in square brackets. In the bottom row, we show the integrated net flux from the 6.9 $\mu$m line in Jy. \label{table} }
\end{center}
\end{table}

\begin{figure*}
    \includegraphics[angle=90,width=\columnwidth]{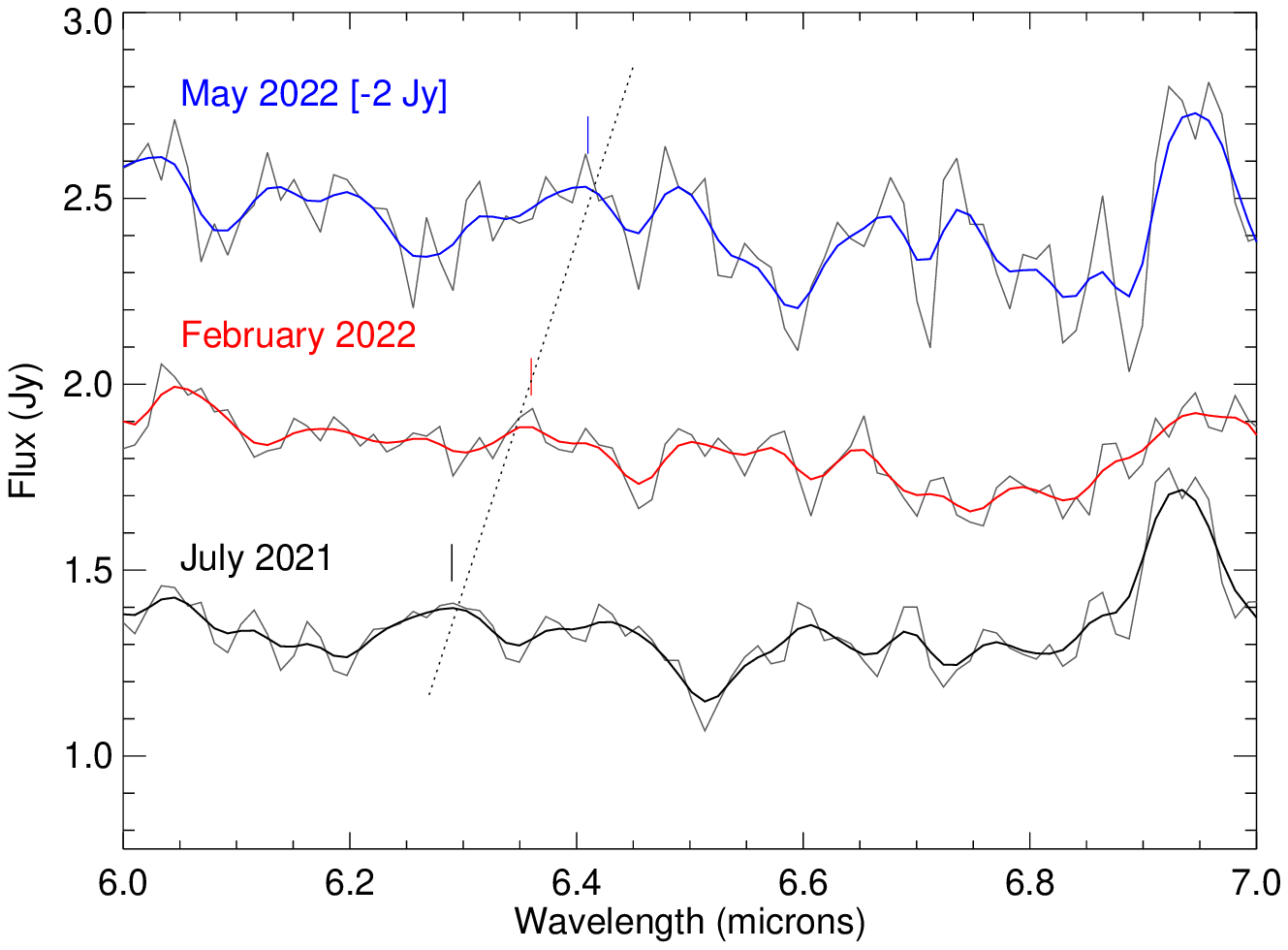}
    \caption{The portion of the SOFIA grism spectroscopy around the UIR feature at 6.2--6.5 $\mu$m. The data from May 2022 were shifted downward by 2 Jy for clarity. The average peak position from our measurements is highlighted and the data are smoothed for clarity with the unsmoothed spectra shown in grey. We have overplotted a dotted line to guide the eye in the shift of the peak wavelength. \label{fig:UIR} }
\end{figure*}

\section{Discussion}
We have found that a seemingly weak emission line at $\sim$6.2-6.5 $\mu$m (see Fig.~\ref{fig:UIR}) in the infrared spectra of WR 137 taken with SOFIA during its last year of operation. The emission line increased in strength as the dust formation began leading up to the periastron passage that will occur in $\sim$2024. \citet{2002ApJ...579L..91C} observed a similar feature in the dusty WR binary WR\,48a. In this discussion, we will compare this feature with that seen in other WCd binaries and how its form changed throughout these three observations and what that could imply for dust creation in WC binaries, and in particular for the case of WR\,137.

\citet{2022ApJ...930..116E} observed a broad 8 $\mu$m feature in the spectrum of the 28.1-year period system WR 125 when the system was near its infrared maximum. This feature was seen in ISO spectra of several WCd stars and was compared to these other systems by Endo et al. The broad 8-$\mu$m feature is often associated with the unidentified infrared (UIR) features, and thus it probably represents emission from the building blocks of the dust particles in the colliding wind region. Our SOFIA spectra of WR\,137 are cut off by the detector and do not cover the region of the 8-$\mu$m feature. 

However, the 8 $\mu$m feature is usually associated with a feature at 6.2 $\mu$m in infrared dust spectra. \citet{1981MNRAS.196..269D} found that the UIR features are often inherent in compounds made of aromatic C$-$H and C$=$C chemical bonds, with some suggesting that this arises from PAH emission \citep[e.g.,][]{1985ApJ...290L..25A}. Since the UIR feature is usually seen at 6.2$\mu$m, the observed wavelength of the feature would imply that the feature is shifted to a longer wavelength in the spectrum of WR\,137. However, it is also seen at a longer wavelength in the spectrum of WR\,48a presented by \citet{2002ApJ...579L..91C}, which is classified as ``class C" according to the procedures outlined by \citet{2002A&A...390.1089P}. 

The longer wavelength of the feature can be caused by hydrogen being absent in the PAH molecules that form this line \citep[e.g.,][]{1997A&A...319..318P}. Furthermore, we saw some evidence that the central  wavelength of the feature shifted to longer wavelengths during the year of our observations. In Fig.~\ref{fig:UIR}, we show the spectra in the region of this line with the average peak wavelength shown. For clarity, we also smooth these spectra and shift the last spectrum to a similar flux level as the other observations for clarity. We caution that the signal-to-noise of these spectra is low as evidenced in the unsmoothed spectra in Fig.~\ref{fig:UIR}, but the overall shift of the line peak went from $\sim$6.29$\mu$m in July 2021 to $\sim$6.35$\mu$m in February 2022, and then to $\sim$6.41$\mu$m in May 2022. We speculate that this shift implies that the dust begins forming in the WR\,137 system in a mixing environment of the WC wind and the Oe hydrogen disk. As periastron nears, the dust is sent along the collision region which becomes more intense. These initial grains can then grow with additional carbon from the WC wind but with less hydrogen available from the Oe wind.

Lau et al.~(submitted) presented aperture masking interferometry of the WR\,137 system at 3.8 $\mu$m and 4.8 $\mu$m taken with the NIRISS instrument on {\it JWST}. The resulting images of the dust around WR\,137 show a very narrow extension of the dust compared to the models that are normally able to reproduce the geometry of the WCd binary dust \citep[e.g., WR\,112,][]{2020ApJ...900..190L}. These models have been widely used both for systems with well-established orbits like WR\,140 \citep{2022NatAs...6.1308L} as well as for inferring orbital information from long-period systems like Apep \citep[WC8+WN;][]{2020MNRAS.495.3323C}, WR 48a \citep[WC6;][]{2020ApJ...898...74L}, or WR 112 \citep[WC8;][]{2020ApJ...900..190L}. The narrow signature of dust formation in the WR\,137 system could imply that the mixing with the Oe star's disk material is important and crucial for the dust formation in this system, which allows for periodic dust formation as the disk is remarkably stable \citep{2020MNRAS.497.4448S}. 

%In the WR 137 system, St-Louis et al. (YYYY) recently showed that the companion O star is actually an Oe star with a remarkably stable disk, showing little or no variability over the course of several months of regular observations. This means that the O star has a stable equatorial decretion disk similar to those around the cooler and less massive Be stars. We speculate that the combination of the AMI observations presented by Lau et al. (2023) and the changing central wavelength of the 6.2-6.4 um feature we observe can allow us to speculate on the formation of the dust in this system. 

We suggest that the wind of the WC star collides with the equatorial disk surrounding the Oe star. As this begins, the disk provides a large reservoir of hydrogen that mixes with the carbon-rich WC wind. This mixing begins at times where the PAH-like formation can create molecules that are not very hydrogen-deficient. However, as the collisions in the system build and the binary approaches periastron, we see the feature move to longer wavelengths indicating a potential for fewer hydrogen atoms to be bound to the PAHs emitting from the feature. As a result of this mixing, the dust is built from the disk, but gradually grows with more carbon-rich material than with the mixed materials. 

The infrared light curve of WR\,137 (Fig.~\ref{fig:LC}) has been seen in the past to be somewhat more irregular than other episodic dust-producing WC binaries. In the WR 140 system, the light curve is almost completely repeatable with very few or no excursions. However, in the light curve of WR\,137, \citet{2001MNRAS.324..156W} show evidence of ``mini-eruptions" happening in 1987, 1988, 1990 on top of the regular activity. The current outburst shows some deviations from these previous outbursts. These differences from the repeatable behavior of WR\,140 shows that variable disk activity from the Oe star could be a protagonist in our understanding of dust production for WR\,137.

The observations presented here, especially when coupled to the stellar parameters presented by \citet{2020MNRAS.497.4448S} and the {\it JWST} observations presented by Lau et al.~(2023), provide some speculative evidence for mixing being an important aspect for dust creation in the WCd binaries with episodic dust creation, or at least for the case of WR\,137. Lau et al.~(2023) also presented evidence that the collision strength between the two winds for WR 137 was not strong enough for regular dust formation, stating that the disk was required for the dust formation in WR 137 in order to provide an environment with high enough density to begin the dust formation process. 

The disk seems to be an essential ingredient in this system’s dust creation, so we also want to consider how such a star evolved in the system. Oe stars, like the Be stars \citep{2013A&ARv..21...69R}, are rapid rotators. This is also observed in WR\,137 where \citet{2016MNRAS.461.4115R} measured a $v \sin i$ of 220 km s$^{-1}$ for the O star, although this is close to an average rotational velocity for a Be star \citep[e.g.,][]{2007A&A...472..577M}. The results of Marchenko et al. (1999?) and Richardson et al.~(2016) seem to point towards an edge-on orbital geometry for WR 137, a result that is being confirmed with follow-up interferometry (Richardson et al., in prep), which could mean the Oe star is not aligned with the orbit.

\section{Conclusions}

We have found that the mid-infrared emission line at 6.3--6.4 $\mu$m in the spectrum of WR\,137 is associated with the growth of dust during the approach towards periastron passage in 2021--2022. The feature grew in strength as the continuum emission from the dust also grew while the relative strength of the emission lines from the Wolf-Rayet wind appeared to shrink due to the increasing continuum. The peak wavelength of the feature appeared to shift to redder wavelengths as the feature grew, likely associated with hydrogen being less prevalent in the grains as the grains grew. We do caution that this is a marginal detection with the signal-to-noise of the data, but the long dust plume, as recently imaged with aperture-masking interferometry with {\it JWST}, is incredibly narrow. We speculate that these results show that the dust grain growth begins in a mixing of material from the WC wind and the Oe disk. Such a mixing is less important as the dust continues to build, leading to a shift in the peak wavelength of the 6.2 $\mu$m feature that is typically considered a UIR feature. 

SOFIA observed both WR\,137 and WR\,125 (Daly et al., in prep) during recent dust-producing episodes. These data along with some infrared data taken from the ground \citep[e.g.,][]{2022ApJ...930..116E} provide a look into the dust formation at critical epochs in the binary clocks for these systems. Such observations will provide a pivotal context for interpretation of higher-resolution and higher signal-to-noise spectroscopy of spatially resolved dust around these massive binaries as has already been accomplished with \textit{JWST} with the prototype of these systems, WR\,140 \citep{2022NatAs...6.1308L}.

%\begin{acknowledgments}

Based on observations made with the NASA/DLR Stratospheric Observatory for Infrared Astronomy (SOFIA). SOFIA was jointly operated by the Universities Space Research Association, Inc. (USRA), under NASA contract NNA17BF53C, and the Deutsches SOFIA Institut (DSI) under DLR contract 50 OK 2002 to the University of Stuttgart. Financial support for this work was provided in part by NASA through award \#09-0163 issued by USRA. MJP is grateful to Embry-Riddle Aeronautical University's Undergraduate Research Institute for financial support through their IGNITE program. NDR is grateful for support from the Cottrell Scholar Award \#CS-CSA-2023-143 sponsored by the Research Corporation for Science Advancement.

%\end{acknowledgments}


\vspace{5mm}
\facilities{SOFIA(FORCAST)}


\software{astropy \citep{2013A&A...558A..33A,2018AJ....156..123A}
          }



% \appendix

% \section{Appendix information}


\bibliography{sample631}{}
\bibliographystyle{aasjournal}



\end{document}